# Modeling the Probability of Failure on LDAP Binding Operations in Iplanet Web Proxy 3.6 Server

Alejandro Chinea Manrique de Lara
Alejandro.Chinea@sun.com

**Abstract.** This paper is devoted to the theoretical analysis of a problem derived from interaction between two Iplanet products: Web Proxy Server and the Directory Server. In particular, a probabilistic and stochastic-approximation model is proposed to minimize the occurrence of LDAP connection failures in Iplanet Web Proxy 3.6 Server. The proposed model serves not only to provide a parameterization of the aforementioned phenomena, but also to provide meaningful insights illustrating and supporting these theoretical results. In addition, we shall also address practical considerations when estimating the parameters of the proposed model from experimental data. Finally, we shall provide some interesting results from real-world data collected from our customers.

## I. INTRODUCTION

In release 3.52 of Iplanet Web Proxy Server [1] support for native LDAP services was included in addition to the standard features provided by the Proxy server. The main goal, between others, was to allow a massive storage of user profiles for authentication purposes. The principal advantage was to overcome the traditional problem of the limited capacity of the local database. However, together with the improvement, additional problems will appear. More specifically, derived from the way LDAP binding connections are made by the proxy for authentication purposes, as well as, the management made by the Directory server regarding binding operations. In this article we propose a methodology capable of dealing with such kind of irregular behaviour, as we shall shortly demonstrate.

The rest of this paper is organized as follows: In the next section we shall present more in detail aspects of the internal architecture of Iplanet Web Proxy 3.6 Server. At this point, we shall address the above-mentioned problem derived of the way LDAP connections from the Proxy to the Directory server were structured. Section III is devoted to the development of a theoretical model in order to characterize the aforementioned phenomena. As we shall see, the proposed model will permit us to minimize the impact of such undesirable behavior. Afterwards, in Section IV we shall address the problem of estimation of the parameters of the proposed model through a stochastic approximation procedure. In section V we shall see some experimental results when applying the proposed model. Finally, section VI we shall outline some concluding remarks to the present study.

## II. PROBLEM STATEMENT

The internal architecture of Iplanet Web Proxy 3.6 Server was structured as a multi-process architecture. At Proxy start-up, a master process or daemon has the responsibility, firstly, of creating the listen socket in charge of accepting connections, secondly, of forking as many child processes as indicated by the configuration files. It's important to note that inside processes there are no worker threads. In particular, each process serves only one request each time.

One of the features provided by the Iplanet Web Proxy 3.6 server is its ability to restrict the access of the data being served (eg: specific URL's). More specifically, we can restrict the access to only certain pre-defined people to different resources like URL's or protocols. The process to configure access control within the proxy starts by choosing either an LDAP Directory Server [2] either a local database to enter the users we want to restrict or give access. Afterwards, when the proxy server evaluates an incoming request determines access based on a hierarchy of rules called access-control entries (ACE's), and then uses the matching entries to determine if the request is allowed or denied. In addition, if SSL is enabled in the proxy, certificate authentication can be also used in conjunction with access control.

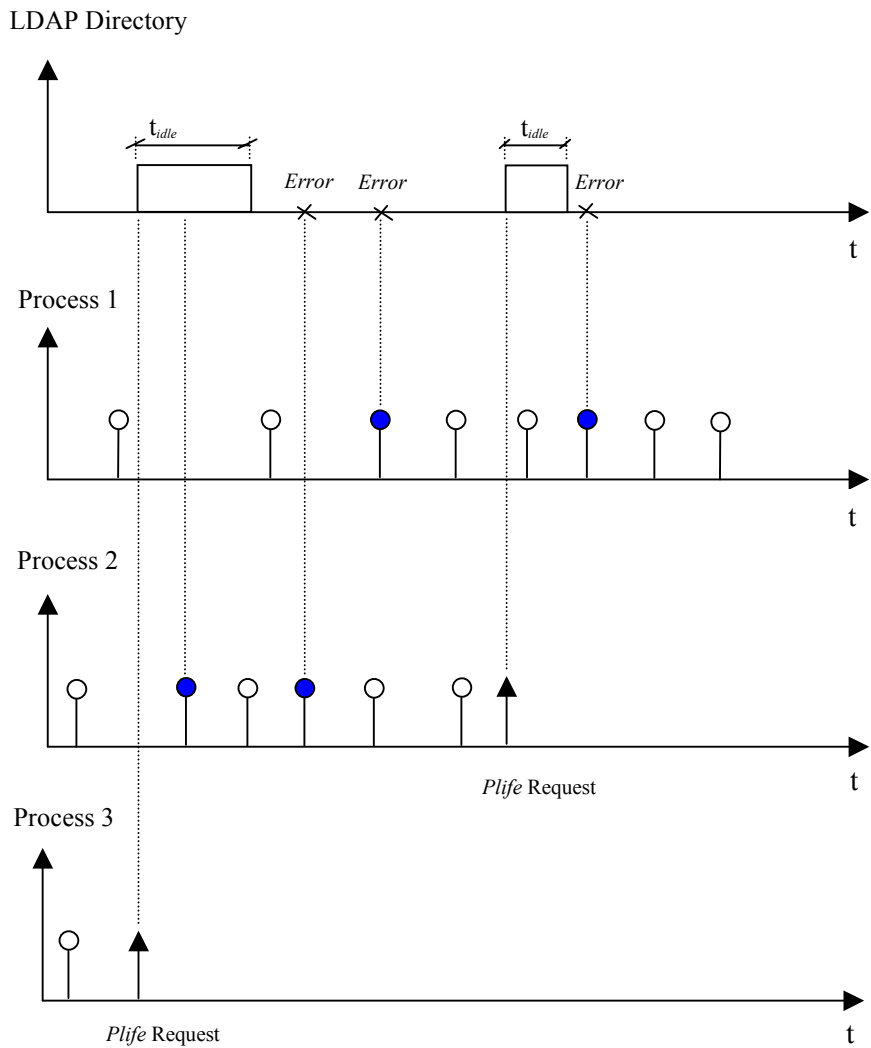

**Figure 1**. Schematic representation of the problem. The incoming requests served by the processes are depicted as a line with a circle on top.

In all of the above cases, the process of authentication will be linked to the fact of establish one or more connections to the LDAP Directory server to check the credentials provided by the users against the information previously stored in the Directory or local database. In order to accomplish such a goal, at proxy start-up the daemon process creates a pool of connections against the Directory server or the local database. Specifically, it creates as many connections as indicated by the magnus.conf file variable *LDAPConnPool*. In addition, the daemon process implements a mobility rule which periodically calls a function in charge of checking the sanity of the LDAP connection pool.

The period of function callbacks is determined by the *ProcessLife* configuration parameter, which indicates, the number of requests each child process serves during its lifetime. Once any of the child processes has served *ProcessLife* request exits throwing a SIGCHLD signal which is trapped by the daemon process which respawn a new process, checking at the same time the status of the LDAP connection pool.

However, from the point of view of the Directory server, there is a parameter which controls the time a connection can stay in idle state, that is, without no activity. More specifically, in Iplanet Directory server 5.x this parameter is called *nsslapd_idletimeout*. Therefore, if one of the connections stay idle more than the time indicated by the *nsslapd_idletimeout* parameter the LDAP Directory server will dropp the connection. Hence, if one of the child process tries to use the connection will get an error which will be translated later on, into an authentication error.

The aforementioned problem is schematically depicted in figure 1. The requests served by the processes are represented as a line with a circle on top. From the incoming request not all of them will lead to a connection to the Directory server, for instance, because we have enabled LDAP Caching or simply because we are accessing a non restricted resource. This fact, is represented by the color of the top circles of the requests, the white ones correspond to connectionless request. In addition, the last request served by each process is represented by a line with a triangle on top. Whenever, a connection arrives to the Directory the timer for the parameter *nsslapd_idletimeout* is restarted if and only if the connection has not been dropped in the meanwhile. Moreover, when a process serves its *ProcessLife* requests the timer is reinitialized even if the connection has been dropped as a consequence of the checkup made by the daemon process.

Taking this considerations into account, in the next section we shall see how can we model the abovementioned behaviour in order to minimize the probability of a connection failure.

## III. MODELING BINDING OPERATIONS OF WEB PROXY SERVER

The first step me must made in order to model the probability of failure is to characterize the incoming flow of request to the proxy. It can be easily deduced, that the nature of the aforementioned process is inherently stochastic.

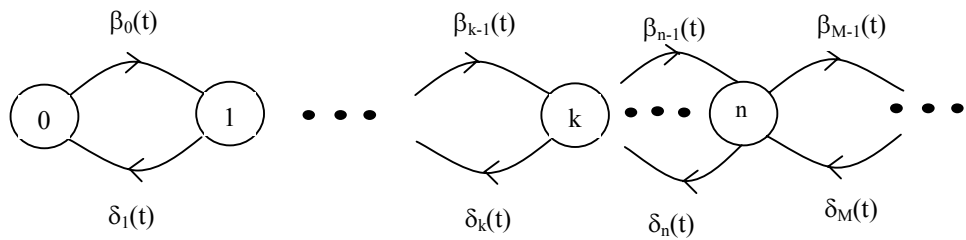

**Figure 2**. Markov chain of a renewal process. For simplicity autoloops are not depicted in the figure.

Specifically, we are considering, as one of the depart hypothesis, that the requests are generated by a finite number of users, that is, we are considering a finite number $N$ of stochastic sources corresponding to the number of users declared within the proxy (the users created trough the proxy admin GUI which are finally LDAP Directory server users). Bearing in mind this considerations, we can model the incoming requests to the proxy as a renewal process, which are a particular class of discontinuous Markov processes. Specifically, they suppose that the possible transitions, supposing the system is in state $S_k$, are limited to state $S_{k+1}$ (a birth) or to state $S_{k-1}$ (a death). This limited amount of memory suffices to produce a great diversity of behaviors.

In order to model the dynamic behavior of such a system we can use the Chapman-Kolmogorov relationship [3], [4] which particularized for a causal system leads to the forward equations. In case of a renewal process the partial stochastic differential equations can be reduced to the expressions:

$$\frac{dP_0(t)}{dt} = -\beta_0(t)P_0(t) + \delta_1(t)P_1(t) \quad k = 0$$

$$\frac{dP_k(t)}{dt} = \beta_{k-1}(t)P_{k-1}(t) - [\beta_k(t) + \delta_k(t)]P_k(t) + \delta_{k+1}(t)P_{k+1}(t) \quad k \geq 1$$

(1)

In the above stochastic differential equations the term $P_k(t)$ represents the probability of finding the system in state k, while the amounts $\beta(t)$ and $\delta(t)$ represents the rates of birth and death which depends of the state considered. The initial condition for the above equations is that at time $t=0$ the number of request is equal to one, that is, $P_0(0) = 1$. In our case, we are observing the activity generated by N users. Therefore, at any instant of time t, we suppose that if the observer is in state $S_k$, that is having observed (registered) k requests from the population, the possible transitions in the interval *[t , t + Δt]* (with *Δt* tending to zero) are to state $S_{k+1}$ ,arrival of a new request as we are simply registering the incoming requests observed from the finite population. That is, the rate of deaths is zero for the whole state space. It's important to note that our objective is simply to characterize the statistic distribution of the stochastic process attacking the proxy.

If we suppose that the users follow the same statistic, which is a reasonable hypothesis given that users tend to perform requests with their browsers more or less at the same interval times (eg: in office working hours). Thus, the probability that a user performs a request in the interval *[t , t + Δt]* conditioned to the fact that in time *t* user was idle will be: *βΔt + o(Δt)*. Therefore, the probability that *i* users from a total of *j* users in idle state at time *t*, perform a request in the interval *[t , t + Δt]* will be given by the binomial distribution:

$$\binom{j}{i}[\beta\Delta t + o(\Delta t)]^i [1 - \beta\Delta t - o(\Delta t)]^{j-i}$$

Therefore, for i = 0 we have 1-β$\Delta$t-o(t), for i =1 jβ$\Delta$t+o(t) and finally for i> 1 o(t). Thus, we can identify the rate of births as $\beta_k$ = (N-k) β. Figure3 depicts the resulting Markov chain:

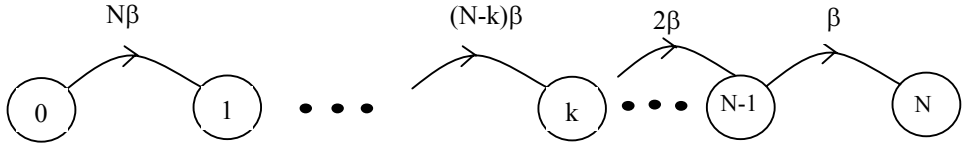

**Figure 3**. Markov chain of the renewal process which models the activity of N proxy users.

Taking into account the considerations stated above, the resolution [5], [6] of differential equations (1) particularized for the above Markov chain lead to the following distribution:

$$P_k(t) = \binom{N}{k} e^{-(N-k)\beta t} \left(1 - e^{-\beta t}\right)^k \quad k \geq 0 \quad (2)$$

Which correspond to a Bernoulli process. At this point, our main objective is to be able of modeling the probability of failure as a consequence of the phenomena explained in section II. Hence, the aforementioned probability of error is equal to the probability of the event $E$ = *"The interval time between two succesive marked request is bigger than nsslapd_idletimeout"*. Meaning with the term "marked" those request from the incoming flow that are translated to an LDAP connection. It's important to note, as we had previously explained, that not all of the incoming requests will be directly translated into a LDAP connection for instance because the caching of LDAP request is active in the proxy. Therefore, we can compute the probability of event $E$ as follows, using the results obtained in (2):

$$P(E) = \sum_{i}^{\infty} P\{P(t) = i\} P\{E / P(t) = i\} \quad (3)$$

Where P(t) is the stochastic process which is attacking the proxy, that is, P{P(t) = k} is given by expression (2) while the second term of the computation is the probability of error conditioned by the fact that a number of request equal to i has been produced in interval [0, t] with t = *nsslapd_idletimeout*. If we denoted by ξ the probability of a marked request, it can be easily deduced that the second term is equivalent to compute the probability of the event *"In interval [0 nsslapd_idletimeout] arrives one non marked request if i =1, arrives two non marked request if i=2 and so forth"*. Of course, in the previous reasonings we are considering as origin of time $t = 0$, and we also suppose that $t = 0-\Delta t$ (with $\Delta t$ tending to zero) we have received a marked request. Summing up, substituting in the previous expression we obtain:

$$P(E) = \sum_{i}^{N} \binom{N}{i} e^{-(N-i)\beta t} (1-\xi)^{i} (1-e^{-\beta t})^{i}$$

The above result admits the following closed expression:

$$P(E) = \left(1 - \xi\left(1 - e^{-\beta \text{nsslapd\_idletimeout}}\right)\right)^{N} \quad (4)$$

From the above expression we can observe that results are coherent, at the extent the probability of a marked request decreases the probability of error increases. At this point, we could be tented of minimizing expression (4) with respect to the probability of a marked request $\xi$. However, the minimum (which is a zero of the probability function) is achieved for a value of $\xi > 1$ which is not feasible for a probability function. Furthermore, finding a closed expression for the probability $\xi$ is very complex as depends of a lot of factors: configuration parameters of LDAP Cache, the value of the parameter *ProcessLife* and so forth. However, as we shall see in the next section we can estimate its approximate value empirically. Therefore, from a practical point of view we shall play with two parameters: *nsslapd_idletimeout* and the probability of error $P(E)$ (hereafter $\varepsilon$ ). Re-arranging the previous expression we obtain:

$$nsslapd\_idletimeout = -\frac{1}{\beta} \log\left[1 - \frac{1 - e^{\frac{\log \varepsilon}{N}}}{\xi}\right] \quad (5)$$

Specifically, fixing the probability of error $\varepsilon$ we want to achieve, we can compute the value of the parameter *nsslapd_idletimeout*. Note, that the expression provide us with precious information regarding a practical limit on the probability of error we can expect to achieve. Note that the right operand of the logarithmic expression must be inside the interval [0,1], hence when fixing the probability of error, the following lower bound must be respected:

$$\varepsilon > (1-\xi)^{N}$$

Bearing in mind this fact we can extract the following conclusions, which validate what we could intuitively expect:

- At the extent, the population of users $N$ increase the lower bound for the probability of error $\varepsilon$ decrease. Furthermore, it would be the dominant factor in expression (5) leading to lower values for the timeout needed. Moreover, by using a series expansion, expression (5) can be approximated for $N$ big by:

$$nsslapd\_idletimeout \approx -\frac{1}{\beta\xi}\frac{\log\varepsilon}{N} \quad (6)$$

- If the activity of the users, represented by the parameter *β* increase, for a fixed probability of error *ε* the time value *nsslapd_idletimeout* needed decrease.

- If the probability of a marked request *ξ* increases the probability of error ε decreases. Furthermore, fixing the probability of error to a value the fact of increasing *ξ* (for instance, by lowering the magnus.conf variable *ProcessLife*) is translated on a reduction of the value needed for *nsslapd_idletimeout*.

In the next section we shall address the estimation of the parameters of the proposed theoretical model.

## IV. APPLICATION OF AN STOCHASTIC APPROXIMATION ALGORITHM TO COMPUTE MODEL PARAMETERS

Probability theory provides a mathematical framework for the study of random phenomena. It requires a precise description of the outcome of an observation when such a phenomenon is observed. In this line, the probability *P(A)* of an event *A* measures the likeliness of its occurrence. Hence, in a first approach, we could approximate the probability of a marked request as the empirical frequency of occurrence of such an event in *n* independent experiments. Let's denote by *S* the event *"Arrival of a marked request"*, therefore, we could approximate *P(S)* as follows:

$$P(S) \approx lim_{n->\infty} \frac{n_s}{n}$$

Specifically, if *n* independent experiments are performed, that is, the processing of n request by the proxy, among $n_s$ will result in the realization of event *S*, meaning that from the requests processed by the proxy which of them were translated to a connection to the Directory Server. Thus, from the access log we can compute the total number of processed request by the proxy in a specific interval time. In addition, from the error log we can compute the number of connections established against the LDAP Directory Server as in such a file whenever a connection to the Directory server is made for authentication purposes, proxy logs any of the accesses. Moreover, from the access log of the Directory Server we could also compute the number of binding operations coming from the Proxy by using an appropriate script program.

However, such a procedure has several drawbacks, firstly, we are implicitly supposing the ergodicity of the underlying process , that is, the existence of an stationary regimen which is not always true. Furthermore, we cannot track the variations on the statistic of users neither on the average value of probability $\xi$. Thus, we must provide a robust mechanism capable to deal with such a problem. In order to accomplish such goal we propose the following recursive stochastic approximation algorithm:

**1.-** Choose an averaging period T, and fix the desired probability of error $\varepsilon$

**2.-** Initialize $\xi_0 = 0$, $\beta_0 = 0$. Obtain first estimation of the probability of a marked transition and user statistic $\chi_0$ and $\theta_0$ respectively using the empirical frequency of events over the averaging period T.

**3.-** Compute the recursion:

$$\begin{bmatrix} \xi_{n+1} \\ \beta_{n+1} \end{bmatrix} = \begin{bmatrix} \xi_n \\ \beta_n \end{bmatrix} + \eta_n \begin{bmatrix} \chi_{n+1} - \xi_n \\ \theta_{n+1} - \beta_n \end{bmatrix}$$

Where $\eta_0$ is a sequence accomplishing the properties:

$$\eta_n > 0 \quad \lim_{n \to \infty} \eta_n = 0 \quad \sum_n^\infty \eta_n = \infty$$

**4.-** Compute *nsslapd_idletimeout* using expression (5) or (6) depending on the value of $N$

$$nsslapd\_idletimeout \approx -\frac{1}{\beta_{n+1}} \log\left[1 - \frac{1 - e^{\frac{\log \varepsilon}{N}}}{\xi_{n+1}}\right]$$

**5.-** If below condition is reached update real parameter in Directory Server:

$$\left| nsslapd\_idletimeout_{current} - nsslapd\_idletimeout_{previous} \right| \geq \Delta$$

**6.-** Goto step 3

The idea behind above procedure is to use a recursive estimator to track the variations [6], [7] of the underlying processes. One takes an observation at the current estimator of the parameters of our model, then uses that observation to make a small correction in the estimate, then takes an observation at the new value of the estimator and so forth. The fact that the step sizes are small is important for the convergence, because it guarantees an "averaging of the noise". Decreasing step sizes provides an implicit averaging of observations. In our experiments we have chosen for $\eta_n$ the harmonic sequence:

$$\eta_n = \frac{1}{n+1}$$

With regards the parameters of the algorithm, the averaging period T is the time frame to be chosen for performing the computation of the empirical frequency of events. Specifically, we must perform the computation of the statistic of the users $\theta$ and the probability of a marked request $\chi$. The parameter $\Delta$ is just a threshold to notice of a strong variation on the timeout parameter which would be updated in that case.

## V. PRACTICAL RESULTS

This section presents the results obtained by applying the previous algorithm with experimental data collected from Iplanet customers. In the experiments both expressions, (5) and (6) were used. In particular, expression (6) was used for values of $N > 500$. About the averaging period $T$ was chosen ranging between 20 minutes to a few hours. We stopped the algorithm either in the first iteration either after 4 or 6 iterations due to a lack of data. However, the purpose was simply to obtain an approximate idea about how the model works.

| *Prob. of error $\varepsilon$* | *$\beta$ (rq/s)* | *$\xi$* | *N* | *nsSlapd_idleTimeout (s)* |
|---|---|---|---|---|
| 0.1 | 8.3E-4 | 0.3947 | 800 | 8.75 |
| 0.1 | 1.39E-3 | 0.1338 | 150 | 87.03 |
| 0.1 | 0.06 | 0.5887 | 10000 | 0.0065 |

**Table 1**. Experimental results.

From the above table, it can be deduced, as we pointed out in section III, that the value obtained for the LDAP timeout is strongly influenced by the size of the population. More specifically, for big values on the number of users the associated timeout decrease. In addition, the default value of *nsslapd_idletimeout* is zero which means connections can stay in an idle state indefinitly. However, in most of the customer environments the default value is changed in order to save resources on Directory server side. Therefore, only in specific cases where the number of users and/or the traffic is very small and the default settings has been altered we should pay attention to the value of *nsslapd_idletimeout* to be used.

In addition, a more detailed study should address the trade-off when selecting the aforementioned timeout in terms of resource consumption. This further study would hopefully permit to optimize network resources depending on the traffic pattern and the value of the timeout selected. Furthermore, it could be extended to other Iplanet products like Web Server, Directory Server (specifically, to the chaining mechanism) or Directory Server Access Router as similar mechanism arises.

## VI. CONCLUSIONS

In this paper, after reviewing some details of the internal architecture of Iplanet Web Proxy 3.6 server, we have addressed a problem motivated by the way LDAP binding operations were structured for authentication purposes.

In order to tackle this undesired behavior, a theoretical model was proposed with the aim of both, minimizing its influence, and assess the impact of such phenomenon in current customer environments. Afterwards, as one of the results of the present study, a recursive stochastic approximation algorithm was developed to determine the parameters of the proposed model.

In conclusion, it was found that customer environments are affected by the phenomenon described through this paper only when the number of users and/or the traffic is very small. As a rule of thumb setting the parameter *nsslapd_idletimeout* around 60 seconds is quite enough for minimizing the impact of connection failures in most of the cases. Furthermore, we have provided with a robust procedure to follow which is able to provide qualitative and quantitative answers to the problematic we have addressed through this paper. Future research should be focused on the way the selection of the timeout affects to resource consumption in order to develop an adaptive algorithm capable of optimizing resources no matter traffic conditions are. Finally, the proposed methodology also provides with precious insights for system design in other current Iplanet products like Directory Server, Web Server or Directory Server Access Router.